\documentclass[aps,prb,preprint,superscriptaddress,showpacs,preprintnumbers,amsmath,amssymb,floatfix]{revtex4-1}
\usepackage{graphicx}
\begin{document}
\preprint{}

\title{Dissipative-coupling-assisted laser cooling: Limitations and perspectives}

\author{Alexander K. Tagantsev}
\email{alexander.tagantsev@epfl.ch}
\affiliation{Swiss Federal Institute of Technology (EPFL), School of Engineering, Institute of Materials Science, CH-1015 Lausanne, Switzerland}
\affiliation{Ioffe Phys.-Tech. Institute, 26 Politekhnicheskaya, 194021, St.-Petersburg, Russia}

\begin{abstract}
The recently identified possibility of ground-state cooling of a mechanical oscillator in the unresolved sideband regime by combination of the dissipative and dispersive optomechanical coupling under the red sideband excitation [Phys. Rev. A 88, 023850 (2013)], is currently viewed as a remarkable finding.
We present a comprehensive analysis of this protocol, which reveals  its very high sensitivity to small imperfections such as an additional dissipation, the inaccuracy of the optimized experimental  settings, and the inaccuracy of the theoretical framework adopted.
The impact of these imperfections on the cooling limit is quantitatively assessed.
A very strong effect on the cooling limit is found from the internal cavity decay rate which even being small compared with the detection rate may drastically push that limit up, questioning the possibility of the ground state cooling.
Specifically, the internal loss can only be neglected if the ratio of the internal decay rate to the detection rate is much smaller than the ratio of the cooling limit predicted by the protocol to the common dispersive-coupling assisted sideband cooling limit.
More over, we establish that the condition of applicability of theory of that protocol is the requirement that the latter ratio is much smaller than one.
A detailed comparison of  the cooling protocol in question  with the dispersive-coupling-assisted protocols which use the red sideband excitation or feedback is presented.

\end{abstract}

\date{\today}
\maketitle
\newpage

\section{Introduction}
During the past decade the dissipative optomechanical coupling introduced into optomechanics  by Elste, Girvin, and Clerk \cite{Elste2009}
attracted an appreciable attention of theorists~\cite{huang2017,Weiss2013,Weiss2013a,Kilda2016,
vyatchanin2016,nazmiev2019,Vostrosablin2014,Tarabrin2013,Xuereb2011,Tagantsev2018,tagantsev2019,mehmood2019,Khalili2016,huang2018,huang2018quad,huang2019gen,mehmood2018}
and experimentalists~\cite{Li2009,Sawadsky2015,tsvirkun2015,Wu2014,meyer2016,zhang_2014}.
For such a coupling, in contrast to that dispersive, the mechanical oscillator modulates the decay rate of the cavity but not its resonance frequency.
The dissipative coupling has brought about some new physics in optomechanics.
For example, once this coupling is involved, the theory predicts: a generation of a stable optical-spring effect, which is not-feedback-assited~\cite{nazmiev2019}, a virtually full squeezing of the optical noise, in a system exhibiting no optomechanical instability\cite{tagantsev2019}, and not-feedback-assisted cooling of a mechanical oscillator under the resonance excitation~\cite{Tarabrin2013}.
Here the latter was also documented experimentally~\cite{Sawadsky2015}.

Among the predictions for the dissipative-coupling-based systems the most promising is that on a very efficient laser cooling~\cite{Elste2009,Weiss2013a}.
It is a phenomenon of the weak-coupling regime~\cite{marquardt2007} where the light-pressure-induced contribution to the mechanical damping $\gamma_{\mathrm{opt}}$ is much smaller than the cavity decay rate $\gamma$.
In this regime for an appreciable cooling, the phonon number can be viewed as originated from two contributions: one is due to the quantum noise in the bandwidth of the oscillator and the other is due to that in the bandwidth of the optical cavity.
The former scales as $1/\gamma_{\mathrm{opt}}$, it usually dominates the cooling while the later, scaling as $1/\gamma$, can typically be neglected.
In the system where both dispersive and dissipative coupling are active and under a proper detuning, due to interference effects the first contribution "accidentally" vanishes\cite{Elste2009,Weiss2013a}.
As a result the second "small" term dominates the story, leading to a record-low cooling limit as was theoretically demonstrated by Weiss and Nunnenkamp \cite{Weiss2013a}.
However, once the system is not ideal, e.g., because of  the presence of some internal cavity loss,  such a limit will be pushed up~\cite{Elste2009,Weiss2013a}.
The same holds for the inaccuracy of the optimized detuning $\Delta$.
Keeping in mind the situation where the otherwise leading term "accidentally" vanishes, one expects these nonideality effects to be anomalously strong.
We mean that, at  $\gamma_{\mathrm{int}}/\gamma\ll 1$ or/and $\delta\Delta/\Delta\ll 1$  (here $\delta\Delta$ is for the deviation of $\Delta$ from its optimal value and $\gamma_{\mathrm{int}}$ is the internal decay rate of the cavity), the idealized cooling limit may be substantially affected.
On the same lines, one may be concerned about the impact of inaccuracy of the single-mode Langevin equation used for the calculations~\cite{Elste2009,Weiss2013a}.
The point is that, in terms of more precise calculations, the contribution in question may stay nonzero at any settings.
There also exists an additional limitation for the applicability of the results by Weiss and Nunnenkamp \cite{Weiss2013a}: when these are applied one should check that (i) it is  the weak-coupling regime  and (ii) the cold friction does not make the mechanical oscillator overdamped.

From the above it becomes clear that the experimental implementation of the promising result by Weiss and Nunnenkamp \cite{Weiss2013a}, not speaking about practical technical issues, may be more demanding than just the fulfillment of the optimized settings found in Refs.~\citenum{Elste2009,Weiss2013a}.
This justifies the need to specify the range of applicability of this result and  formulate additional conditions for its practical implementation.
This job is the main subject of the present paper, which is organized as follows.
In Sec.~\ref{WN}, the result  by Weiss and Nunnenkamp is reproduced, presented in a simple form, and  an explicit criterion  for its applicability is given.
In Sec.~\ref{IL}, the impact of the internal cavity loss is evaluated.
Section~\ref{Optimal} is devoted to the impact of the inaccuracy of the optimal settings.
In Sec.~\ref{Beyond}, effects beyond the single-mode Langevin-equation accuracy are addressed.
Section~\ref{comparison} discusses the dissipative-coupling-assisted  protocol  versus those dispersive-coupling-assisted.
Section~\ref{Conclusions} gives a brief resume of the paper.
\section{The result by Weiss and Nunnenkamp  and criterion for its applicability}
\label{WN}
A one-sided optomechanical cavity enabled with the dispersive and dissipative optomechanical couplings is considered,
the  coupling constants being denoted as  $g_\omega$ and $g_\gamma$, respectively.
The system is pumped with a strong monochromatic light (the frequency -$\omega_L$, the photon-flux-normalized complex amplitude - $A_0$).
The fluctuations of the cavity field are described with  the photon ladder Bose operator $\mathbf{a }$ while
the fluctuations of the mechanical variable are described  with  the phonon ladder Bose operator $\mathbf{b }$.
These operators satisfy the following equations~\cite{Elste2009}
\begin{equation}
\label{alin}
\frac{\partial \textbf{a}}{\partial t}+\{\gamma/2-i\Delta\}\textbf{a}
=\sqrt{\gamma}\textbf{A}_{\textrm{in}}+\left[ig_\omega a_0+g_\gamma (a_0-A_0/\sqrt{\gamma})\right](\textbf{b}^\dag+\textbf{b}),
\qquad a_0 = \sqrt{\gamma}A_0/(\gamma/2-i\Delta),
\end{equation}
\begin{equation}
\label{blin}
\frac{\partial \textbf{b}}{\partial t}+\left(\frac{\gamma_{\textrm{m}}}{2}+i\omega_{\textrm{m}}\right)\textbf{b}
=\sqrt{\gamma_{\textrm{m}}}\textbf{b}_{\textrm{in}}+i\frac{x_{\textrm{zpf}}}{\hbar}\textbf{F},
\qquad x_{\textrm{zpf}}=\sqrt{\frac{\hbar}{2m\omega_{\textrm{m}}}},
\end{equation}
where $\Delta=\omega_L -\omega_c$ is the detuning and the operator of the backaction force  has the following form
\begin{equation}
\label{F}
\frac{x_{\textrm{zpf}}}{\hbar}\textbf{F}= g_\omega a_0^*\mathbf{a}+i\frac{g_\gamma}{\sqrt{\gamma}} [(a_0^*\textbf{A}_{\textrm{in}}-A_0^*\textbf{a})] +\textrm{H.c.},
\end{equation}
where $\hbar$  is the Planck constant, $\omega_{c}$ and  $\gamma$ are the resonance frequency and  the decay  rate of the cavity while $m$, $\omega_{\textrm{m}}$, and $\gamma_m$ are the effective mass, resonance frequency and decay rate  of the mechanical oscillator, respectively.
Here $\textrm{H.c.}$ stands for Hermitian conjugated.
Operator $\textbf{A}_{\textrm{in}}$ describes the vacuum noise:
\begin{equation}
\label{Aa}
\begin{array}{cc}
[\textbf{A}_{\textrm{in}}(t),\textbf{A}_{\textrm{in}}^\dag(t')]=\delta(t-t'), \qquad [\textbf{A}_{\textrm{in}}(t),\textbf{A}_{\textrm{in}}(t')]=0, \\
<\textbf{A}_{\textrm{in}}(t)\textbf{A}_{\textrm{in}}(t')> =<\textbf{A}_{\textrm{in}}^\dag(t)\textbf{A}_{\textrm{in}}(t')>=0, \\
  \end{array}
\end{equation}
while $\textbf{b}_{\textrm{in}}$ describes the mechanical thermal noise ($n_{\mathrm{th}}$ stands from the number of thermally excited phonons)
\begin{equation}
\label{bin}
\begin{array}{cc}
[\textbf{b}_{\textrm{in}}(t),\textbf{b}_{\textrm{in}}^\dag(t')]=\delta(t-t'), \qquad [\textbf{b}_{\textrm{in}}(t),\textbf{b}_{\textrm{in}}(t')]=0, \\
<\textbf{b}_{\textrm{in}}(t)\textbf{b}_{\textrm{in}}(t')> =0, \qquad <\textbf{b}_{\textrm{in}}^\dag(t)\textbf{b}_{\textrm{in}}(t')>=n_{\mathrm{th}}\delta(t-t'), \\
  \end{array}
\end{equation}
with $<...>$ and $[...,...]$ denoting the ensemble averaging and the commutator, respectively.

The goal is to find the phonon occupation number.
This is a linear problem, which, in the Fourier domain, can be solved exactly~\cite{Weiss2013,Weiss2013a}.
However, according to Ref. \citenum{Weiss2013a}, an approximate solution, keeping a fair accuracy, provides informative analytical results.

The approximate procedure is as follows.
In the Fourier domain, (\ref{alin}) can be solved with respect to $\mathbf{a}$.
Inserting $\mathbf{a}$ into (\ref{blin}), its $\mathbf{b}$-dependent part leads to a renormalization of the mechanical susceptibility, which can be written as follows
\begin{equation}
\label{sus}
\chi(\omega) = \frac{1}{\Gamma_M(\omega)/2-i[\omega-\Omega_M(\omega)]}.
\end{equation}
The other part yields the stochastic backaction force, $\mathbf{F}_{\mathrm{sb}}(t)$.
If we neglect frequency dependent renormalization of $\gamma$ and $\Delta$ due to the optomechanical coupling, the spectral power density of $\mathbf{F}_{\mathrm{sb}}(t)$, which is defined as
\begin{equation}
\label{SF}
S_{FF}(\omega) = \int dt e^{i\omega t}<\mathbf{F}(t)\mathbf{F}(0)>,
\end{equation}
reads~\cite{Elste2009}
\begin{equation}
\label{SF1}
S_{FF}(\omega) = \frac{|a_0|^2 g_\gamma^2}{\gamma(x_{\textrm{zpf}}/\hbar)^2} \frac{(\omega+\omega_h)^2}{(\gamma/2)^2+(\omega+\Delta)^2}.
\end{equation}
where
\begin{equation}
\label{omh}
\omega_h\equiv 2\Delta +\gamma g_\omega/g_\gamma.
\end{equation}

The mechanical spectrum, which is defined as
\begin{equation}
\label{Sb}
S_{bb}(\omega) = \int dt e^{i\omega t}<\mathbf{b}^\dag(t)\mathbf{b}(0)>,
\end{equation}
can be expressed in terms of $S_{FF}(\omega)$ and  $\chi(\omega)$ as follows~\cite{Weiss2013a}
\begin{equation}
\label{Sb1}
S_{bb}(\omega) = |\chi(-\omega)|^2[\gamma_mn_{\mathrm{th}}+(x_{\textrm{zpf}}/\hbar)^2 S_{FF}(\omega) ].
\end{equation}
The relation
\begin{equation}
\label{n}
n = <\mathbf{b}^\dag \mathbf{b}>=\int S_{bb}(\omega)d\omega/2\pi
\end{equation}
can be used to find the number of phonons in the system, which is denoted as $n$.

Using explicit expressions for $\Gamma_M(\omega)$ and $\Omega_M(\omega)$ as well as Eqs.~(\ref{sus}), (\ref{Sb1}), (\ref{SF1}) and (\ref{n}), one can numerically evaluate the cooling of the mechanical oscillator.
Commonly, to advance analytically, in the expression for $\chi(\omega)$ , one replaces\cite{footnote12} $\Omega_M(\omega)$ with $\omega_M$, which satisfy the equation $\Omega_M(\omega)=\omega$  while $\Gamma_M(\omega)$ is replaced with $\gamma_M=\Gamma_M(\omega_M)$.

In this approximations~\cite{Weiss2013a}
\begin{equation}
\label{n1}
n = \frac{\gamma_m}{\gamma_M}n_{\mathrm{th}}+ \frac{|a_0|^2 g_\gamma^2\gamma^{-1}}{(\gamma+\gamma_M)^2/4+(\omega_M-\Delta)^2}
\left[\frac{(\omega_h -\omega_M)^2}{\gamma_M}+
\frac{(\omega_h-\Delta)^2}{\gamma}+\frac{\gamma+\gamma_M}{4}\right].
\end{equation}

This way calculated $\gamma_M$ can also be obtained using  the following result of the quantum noise approach for the light-pressure-induced mechanical decay rate~\cite{marquardt2007}
\begin{equation}
\label{GM}
\gamma_{\mathrm{opt}}\equiv\gamma_M-\gamma_m = (x_{\textrm{zpf}}/\hbar)^2[S_{FF}(\omega_M)-S_{FF}(-\omega_M)].
\end{equation}

The above approximate treatment is valid if the renormalized mechanical oscillator is weakly damped, i.e.
\begin{equation}
\label{Mdamp}
\gamma_M\ll \omega_M,
\end{equation}
while the optomechanical system is in the weak-coupling regime~\cite{marquardt2007} where
\begin{equation}
\label{iequ}
\gamma_{\mathrm{opt}}\ll\gamma,
\end{equation}
which also practically implies
\begin{equation}
\label{iequ1}
\gamma_M\ll\gamma.
\end{equation}
Obviously, the  neglect of the renormalization of $\gamma$ and $\Delta$, crucial for the calculations, is justified only in the weak-coupling regime.
Thus,  Eqs.(\ref{iequ}) and (\ref{Mdamp}) make a creation of the validity for the whole theory.

Equation (\ref{n1}) can be rationalized: the first term in the brackets is the contribution of the quantum noise in the bandwidth of the mechanical oscillator whereas the second and third are conditioned by the noise in the bandwidth of the optical cavity.
In the weak-coupling regime addressed, the first contribution is expected to be dominant unless some special cancelations take place.

In the case of the purely dispersive coupling, i.e., at $g_\gamma\rightarrow0$ and $g_\omega\neq0$, in Eqs.(\ref{n1}), indeed only the first term in the brackets is to be kept.
This leads to a well-known result for the phonon occupation number, which, for the optimal  detuning $\Delta=-\omega_M$, reads
\begin{equation}
\label{n2}
n = \frac{n_{\mathrm{th}}+ n_{\mathrm{disp}}V}
{1+V}, \qquad V\equiv\frac{|a_0|^2 g_\omega^2}{(\gamma/2)^2+4\omega_M^2}\frac{16\omega_M^2}{\gamma\gamma_m},
\end{equation}
where
\begin{equation}
\label{ndisp}
n_{\mathrm{disp}} = \frac{\gamma^2}{16\omega_M^2}.
\end{equation}
is the minimal phonon occupation that can be reached for the dispersive-coupling-assisted sideband cooling~\cite{marquardt2007,wilson2007} under red sideband excitation.

If the both optomechanical couplings are active, there appears the  possibility of breaking through in the minimal phonon occupation number.
Specifically, at $\omega_h =\omega_M $, i.e. at
\begin{equation}
\label{cond}
2\Delta= \omega_M-\gamma g_\omega/g_\gamma,
\end{equation}
the contribution of  the quantum noise in the bandwidth of the mechanical oscillator  vanishes due to the Fano effect~\cite{Elste2009}.
As a result the minimal phonon number is controlled by  the "small" second and third terms in the brackets in Eq.~(\ref{n1}).
For such a detuning, one finds~\cite{Weiss2013a}
\begin{equation}
\label{n3}
n = \frac{\gamma_m}{\gamma_M}n_{\mathrm{th}}+ U,
\end{equation}
where
\begin{equation}
\label{Uoo}
U\equiv |a_0|^2 \frac{g_\gamma^2}{\gamma^2}
\end{equation}
is proportional to the laser power and
\begin{equation}
\label{GM3}
\gamma_M=\gamma_m +  U\gamma_mG,
\qquad G=\frac{G_0}{1+(3\omega_M/\gamma- g_\omega/g_\gamma)^2}\qquad G_0=\frac{16\omega_M^2}{\gamma \gamma_m}.
\end{equation}
Equation (\ref{n3}) can be also rewritten as follows
\begin{equation}
\label{n33}
n = \frac{n_{\mathrm{th}}}{1+GU}+U.
\end{equation}
Minimization of (\ref{n33}) with respect to the intensity of the pumping light yields the following minimal phonon number
\begin{equation}
\label{n4}
n_{\mathrm{diss}} = n_{\mathrm{th}} \left(\frac{2}{\sqrt{Gn_{\mathrm{th}}}}-\frac{1}{Gn_{\mathrm{th}}}\right),
\end{equation}
which is reached at
\begin{equation}
\label{U}
U =U_0\equiv \frac{\sqrt{n_{\mathrm{th}} G}-1}{G}.
\end{equation}
Next, since we are interested in the situation where $n_{\mathrm{diss}} \ll n_{\mathrm{th}}$,  Eqs.(\ref{n4}) and (\ref{U}) can be rewritten as follows
\begin{equation}
\label{n5}
n_{\mathrm{diss}} = 2 \sqrt{\frac{n_{\mathrm{th}}}{G}}
\end{equation}
and
\begin{equation}
\label{U1}
U_0 = \frac{n_{\mathrm{diss}}}{2}.
\end{equation}

Further optimization is possible by manipulating with the ratio of the optomechanical coupling constants~\cite{Weiss2013a}, specifically, by setting
\begin{equation}
\label{cond1}
\gamma g_\omega/g_\gamma=3\omega_M,
\end{equation}
we maximize $G$ up to $G_0$.
Note that (\ref{cond1}) also implies
\begin{equation}
\label{cond2}
\Delta=-\omega_M.
\end{equation}
This brings us to the following minimal phonon number that can be reached in the presence of the dissipative and dispersive coupling
\begin{equation}
\label{ndiss}
n_{\mathrm{diss}} =\frac{1}{2}\sqrt{\frac{n_{\mathrm{th}}}{Q} \frac{\gamma}{\omega_M}}
\end{equation}
where $Q =\omega_M/\gamma_m$ is the quality factor of the decoupled  mechanical oscillator.
Hereafter, referring to this result we will use  "dissipative-coupling-assisted limit" as shorthand.

This cooling limit is reached at the following photon cavity occupation
\begin{equation}
\label{a0}
|a_0|^2= \frac{n_{\mathrm{diss}}}{2}\left(\frac{\gamma}{g_{\gamma}}\right)^2.
\end{equation}

One readily notice that in the bad cavity limit, i.e., at $\gamma\gg\omega_M$, and if the system is dominated by the dissipative coupling, i.e., at $g_\omega/g_\gamma\ll 1$, $G$ is always close $G_0$ such that (\ref{ndiss}) is valid without satisfying condition (\ref{cond1}), while the detuning is different from that given by Eq.(\ref{cond2}).

One can readily find the range of applicability of the cooling limit given by Eq.(\ref{ndiss}).
Combining (\ref{GM}), (\ref{GM3}), (\ref{ndisp}), and (\ref{U1}), one finds
\begin{equation}
\label{gamm/gamm}
\frac{\gamma_{\mathrm{opt}}}{\gamma}= \frac{n_{\mathrm{diss}}}{2n_{\mathrm{disp}}}.
\end{equation}
Thus, for the validity of Eq.(\ref{ndiss}), condition (\ref{iequ}) requires that
\begin{equation}
\label{gamm/gamm1}
n_{\mathrm{diss}} \ll 2n_{\mathrm{disp}}
\end{equation}
while  condition (\ref{Mdamp}) yields
\begin{equation}
\label{gamm/gamm2}
n_{\mathrm{diss}} \ll 2n_{\mathrm{disp}}\frac{\omega_M}{\gamma}.
\end{equation}

In other words, the validity of  the cooling limit predicted in Ref.~\citenum{Weiss2013a} requires that that limit must be  appreciably deeper than the dispersive-coupling-assisted limit for the red sideband excitation (\ref{ndisp}).

\section{Impact of the internal loss}
\label{IL}
The impact of the internal cavity loss on the Fano effect in question was discussed earlier~\cite{Elste2009, Weiss2013a}.
Specifically, in Ref.~\cite{ Weiss2013a}, it was pointed out that, depending on the ratio of $\gamma_{\mathrm{int}}/\gamma$, the quantum noise interference becomes less perfect, and
ultimately, if  $\gamma_{\mathrm{int}}/\gamma\gg 1$, the force spectrum is a Lorentzian.
However, as was stated in the Introduction, in view of the specifics of the system,   one can expect a strong impact of the internal cavity loss on the cooling limit already at $\gamma_{\mathrm{int}}/\gamma\ll 1$.

Let us show this.
The internal loss entails an additional contribution to the spectral power density of the backaction force, which can be approximated as follows~\cite{Weiss2013a}
\begin{equation}
\label{SFdop}
S_{FF,\mathrm{int}}(\omega) =
\frac{U\gamma_{\mathrm{int}}}{(x_{\textrm{zpf}}/\hbar)^2}  \frac{(\gamma/2)^2+(\Delta +\gamma g_\omega/g_\gamma)^2}{(\gamma/2)^2+(\omega+\Delta)^2}.
\end{equation}
To be exact, in this expression, one should replace  $\gamma$  with the total cavity decay rate $\gamma+\gamma_{\mathrm{int}}$.
In what follows, being interested in the situation where  $\gamma\gg\gamma_{\mathrm{int}}$, we will ignore this replacement.

One readily checks that this contribution  leads to a generalization of (\ref{n33}) to find
\begin{equation}
\label{n30}
n = \frac{n_{\mathrm{th}}+HU}{1+GU}+U \qquad H=\frac{\gamma_{\mathrm{int}}}{\gamma_m}\frac{(\gamma/2)^2+(\Delta +\gamma g_\omega/g_\gamma)^2}{(\gamma/2)^2+(\omega_M-\Delta)^2}.
\end{equation}
For the optimized regime given by Eqs.~(\ref{cond1}) and (\ref{cond2}), the contribution of the internal loss to the minimal phonon number via (\ref{n30}) reads
\begin{equation}
\label{n31}
n_{\mathrm{int}} = \frac{H}{G_0}= \frac{\gamma_{\mathrm{int}}}{\gamma}n_{\mathrm{disp}}\beta, \qquad \beta =\frac{(\gamma/2)^2+16\omega_M^2}{(\gamma/2)^2+4\omega_M^2}.
\end{equation}
Next, the requirement $n_{\mathrm{int}}\ll n_{\mathrm{diss}}$ brings us to the conclusion that the impact of the internal loss  can be neglected if
\begin{equation}
\label{crossover}
\frac{\gamma_{\mathrm{int}}}{\gamma} \ll \frac{1}{\beta}\frac{n_{\mathrm{diss}}}{n_{\mathrm{disp}}}=n_{\mathrm{diss}}\frac{8}{\beta}\left(\frac{\omega_M}{\gamma}\right)^2.
\end{equation}
One readily checks that an identical estimate follows for the requirement
\begin{equation}
\label{doubling}
HU_0\ll n_{\mathrm{th}}.
\end{equation}
Using (\ref{gamm/gamm}), Eq.~(\ref{crossover}) can be also rewritten as follows
\begin{equation}
\label{crossover1}
\gamma_{\mathrm{int}} \ll \frac{2}{\beta}\gamma_{\mathrm{opt}}.
\end{equation}

This result implies, that, roughly, to neglect the impact of the internal loss on cooling, the internal loss decay rate should be much smaller than the light-pressure-induced mechanical damping.
Such a requirement is much more demanding than $\gamma_{\mathrm{int}} \ll \gamma$, which one might expect.

\section{Impact of inaccuracy of the optimal settings}
\label{Optimal}
The cooling limit given by  Eq.(\ref{ndiss}) was obtained as a result of three conditions satisfied: (i)~an optimal detuning [Eq.(\ref{cond})],
(ii)~an optimal laser power [ Eq. (\ref{U})], and
(iii)~an optimal ratio of the coupling constants [Eq.(\ref{cond1})].

The impact of the inaccuracy of the optimal detuning  can readily be evaluated by using Eq.~(\ref{n1}) to find that a small deviation of the detuning $\Delta$ from the optimal value of $(\omega_M-\gamma g_\omega/g_\gamma)/2$ by $ \delta\Delta$ will lead to an additional number of phonons
\begin{equation}
\label{nAD}
n_\Delta =  \frac{U\gamma^2}{(\gamma/2)^2+(\omega_M-\Delta)^2}
\frac{4\delta\Delta^2}{\gamma\gamma_M},
\end{equation}
which, for the optimal settings (\ref{cond1}) and (\ref{cond2}), can be rewritten as follows
\begin{equation}
\label{nAD1}
n_\Delta =  \frac{\delta\Delta^2}{\Delta^2}\frac{(\gamma/2)^2}{(\gamma/2)^2+4\omega_M^2}.
\end{equation}
Next, the requirement $n_\Delta \ll n_{\mathrm{diss}}$ brings us to the conclusion that the impact of  inaccuracy of the detuning $\delta\Delta$ on the phonon number can be neglected if
\begin{equation}
\label{cond4}
\frac{\delta\Delta}{\Delta}\ll\sqrt{ n_{\mathrm{diss}}\frac{(\gamma/2)^2+4\omega_M^2}{(\gamma/2)^2}}.
\end{equation}

Equation (\ref{n33}) readily implies that the impact  of the inaccuracy of the optimal laser power on the cooling limit can be neglected if
\begin{equation}
\label{delaU}
\frac{\delta U}{U_0}\ll 1
\end{equation}
where $\delta U$ is a deviation of $U$ from its optimal value $U_0$.

Equations (\ref{n5}) and (\ref{GM3}) enable evaluation of the  increase of $n_{\mathrm{diss}}$ caused by a small violation of condition $\gamma g_\omega/g_\gamma=3\omega_M$, which reads
\begin{equation}
\label{dn}
 n_g = \frac{n_{\mathrm{diss}}}{2} \left(\delta \frac{3 \omega_{\mathrm{M}}}{\gamma}\right)^2
\end{equation}
where  $\delta\equiv(\gamma g_\omega/g_\gamma-3\omega_M )/(3\omega_M)$,
implying that the inaccuracy associated with this condition can be neglected if
\begin{equation}
\label{dn1}
\delta\ll \frac{\sqrt{2}}{3} \frac{\gamma}{\omega_{\mathrm{M}}}.
\end{equation}

Conditions (\ref{cond4}), (\ref{delaU}), and (\ref{dn1}) suggest that, in the unresolved sideband regime, only the requirement from the tuning inaccuracy may be stringent in the case of very deep cooling (at $n_{\mathrm{diss}}\ll1$ ). i.e.  condition $\frac{\delta\Delta}{\Delta}\ll 1$ does not guarantee a negligible correction to the idealized cooling limit.
As for the resolved sideband regime, the requirements for both the coupling constant ratio and  detuning may be demanding.
\section{Beyond the single-mode Langevin equation}
\label{Beyond}
The key element of the theory discussed is the Fano-effect-driven  cancellation of the contribution to the phonon number from the quantum noise in the bandwidth of the mechanical oscillator.
Such a cancellation is the result of the single-mode quantum Langevin-equation approximation.
Evidently, one cannot exclude that, in terms of more precise calculations, this contribution may stay non-zero at any settings.
This issue can be elucidated for the case of the Michelson-Sagnac interferometer\cite{Xuereb2011,Sawadsky2015}, which nowadays is a good candidate for an experimental implementation of the dissipative-coupling assisted  ground-state cooling.
A virtually exact treatment of this system is available~\cite{Tarabrin2013} on the lines of the so-called "input-output relations"~\cite{footnote5} approach~\cite{Buonanno2003,Danilishin2012,Khalili2016}, a method widely employed in the gravitational-wave community.
The result obtained in Ref.~\citenum{Tarabrin2013} for the spectral power density of the stochastic backaction force  in the signal-recycled Michelson-Sagnac interferometer can be rewritten in terms of a one-sided cavity controlled by a common actions of the  dissipative and dispersive coupling (see APPENDIX) to find
\begin{equation}
\label{NSF}
S_{FF}(\omega) = \frac{|a_0|^2 g_\gamma^2}{\gamma(x_{\textrm{zpf}}/\hbar)^2} \frac{(\omega+\omega_h)^2+(\pi\omega_h\omega/\omega_{\mathrm{FSR}})^2}{(\gamma/2)^2+(\omega+\Delta)^2},
\end{equation}
c.f., Eq.(\ref{SF1}), where $\omega_{\mathrm{FSR}}$ is the cavity free spectral range.
With such a modification the condition $\omega_h =\omega_M$ does not lead any more to the cancellation in question.
Thus, beyond the  Langevin-equation approximation, by using (\ref{NSF}) at the optimized settings, we find the following additional contribution to the phonon number
\begin{equation}
\label{nL}
n_\mathrm{L} =\left(\frac{3\pi}{2}\frac{\omega_M}{\omega_{\mathrm{FSR}}}\right)^2\frac{(\gamma/2)^2}{(\gamma/2)^2+4\omega_M^2},
\end{equation}
implying that this contribution can be neglected if
\begin{equation}
\label{Lancond}
\frac{\omega_M}{\omega_{\mathrm{FSR}}}\ll \frac{2}{3\pi}\sqrt{ n_{\mathrm{diss}}\frac{(\gamma/2)^2+4\omega_M^2}{(\gamma/2)^2}}.
\end{equation}
It is seen that this condition may be more stringent than the criterion of applicability of the single-mode Langevin equation $\frac{\omega_M}{\omega_{\mathrm{FSR}}}\ll 1$.
The presence of  $\omega_{\mathrm{FSR}}$ in Eq.~(\ref{nL}) suggests that this contribution may be attributed to the multimode nature of the interferometer.
\section{Comparison with the dispersive-coupling-assisted protocols}
\label{comparison}
\subsection{Sideband cooling}
An important result of Sec.\ref{IL} is that the theory by Weiss and Nunnenkamp \cite{Weiss2013a} predicts the cooling limit that is always lower than that for the dispersive coupling at the red sideband excitation.
This is an exact analytical result, which is consistent with the results of numerical simulations from Ref.\citenum{Weiss2013a}.
However, the application of this conclusion to a real situation should be done with a reservation for the limitations of the applicability  of this theory, which were presented above.
Among these limitations the most stringent is related to the internal cavity loss, which, even being relatively small, i.e. at $\gamma_{\mathrm{int}}\ll\gamma$, can essentially push up the cooling limit (\ref{ndiss}) to the value given by Eq.(\ref{n31}).
At the same time, remarkably, in the regime dominated by the internal loss but at $\gamma_{\mathrm{int}}\ll\gamma$ , the dissipative-coupling-assisted cooling still yields the minimum phonon number a factor of $\beta\gamma_{\mathrm{int}}/\gamma$, with $1<\beta<4$, smaller than the dispersive-coupling-assisted cooling limit.

The cooling limit of a protocol is not its only merit.
The in-cavity photon number needed to approach the limit  also matters.
To characterize the dispersive-coupling-assisted cooling, one can use the phonon number corresponding to the phonon occupancy $2n_{\mathrm{disp}}$, i.e.  twice the
dissipative-coupling-assisted limit.
Using (\ref{n2}), the photon number in question reads
\begin{equation}
\label{a01}
|a_0|^2= \frac{n_{\mathrm{th}}}{Q}\frac{\omega_M}{\gamma}\frac{(\gamma/2)^2+4\omega_M^2}{\gamma^2}\left(\frac{\gamma}{g_{\omega}}\right)^2.
\end{equation}
Equation (\ref{a01}) is to be compared with Eq.(\ref{a0}), which gives the in-cavity photon number  needed to reach the cooling limit (\ref{ndiss}).
To have a reference point, we set $g_\omega \cong g_\gamma$.
For such a setting, comparing (\ref{a01}) with (\ref{a0}) and (\ref{ndiss}) one may conclude that, for typical experimental parameters,  Eq.(\ref{a0}) requires a much larger photon number.
Thus, for the lower dissipative-coupling-assisted  limit, the price of a higher in cavity field has to be paid.
This may question the advantage of the dissipative-coupling-assisted protocol.
However, for a balanced judgment, one can compare  (\ref{a01}) with the in-cavity  photon number needed to reach the level of $2n_{\mathrm{disp}}$ phonon via the other protocol.
Taking into account that $n_{\mathrm{disp}}$ must be much larger than  $n_{\mathrm{diss}}$ and using (\ref{n33}), the aforementioned in-cavity photon number can be evaluated as follows
\begin{equation}
\label{a02}
|a_0|^2\approx \frac{n_{\mathrm{th}}}{2Q}\frac{\omega_M}{\gamma}\left(\frac{\gamma}{g_{\gamma}}\right)^2.
\end{equation}
Comparing (\ref{a01}) with (\ref{a02}), one concludes that, in the sideband resolved regime where the dispersive-coupling-assisted  protocol is commonly viewed as the ultimate tool, the other protocol may require a much a smaller in-cavity photon number for the same cooling level.
For  $g_\omega \cong g_\gamma$, the gain is about $8(\omega_M/\gamma)^2$.

Thus, in many aspects, the dispersive-coupling-assisted protocol looks advantageous for sideband cooling.
\subsection{Feedback-assisted cooling}
As is commonly recognized~\cite{Aspelmeyer2014,Elste2009,Weiss2013a}, the principle advantage of the dissipative-coupling-assisted protocol is the possibility of ground-state cooling in the unresolved sideband regime.
Another cooling protocol that enables  ground-state cooling in that regime is the feedback-assisted cooling via the common dispersive coupling.
Let us compare these protocols.
For the latter, using a well-known result~\cite{rossi2018}, ground-state cooling is possible with a phonon number that can be approximated as follows:
\begin{equation}
\label{fbn}
n_{\mathrm{fb}} =n_{\mathrm{det}} + \frac{4}{\sqrt{\eta_{\mathrm{det}}}}n_{\mathrm{th}}n_{\mathrm{imp}}.
\end{equation}
where $n_{\mathrm{det}}=0.5(\sqrt{1/\eta_{\mathrm{det}}}-1)$ is the detector controlled limit,
\begin{equation}
 n_{\mathrm{imp}}= \frac{\gamma\gamma_m}{64|a_0|^2g_\omega^2}
\end{equation}
is the number of imperfection noise quanta, and $\eta_{\mathrm{det}}$ is the detector efficiency.
Equation (\ref{fbn}) is to be compared with the result by Weiss and Nunnenkamp \cite{Weiss2013a}
\begin{equation}
\label{Mndiss}
n_{\mathrm{diss}} =\frac{1}{2}\sqrt{\frac{n_{\mathrm{th}}}{Q} \frac{\gamma}{\omega_M}}.
\end{equation}

Upon comparing these two cooling  protocols one may notice the $\sqrt{n_{\mathrm{th}}}$-versus-$n_{\mathrm{th}}$ difference between  Eqs.~(\ref{Mndiss}) and (\ref{fbn}) makes the dissipative-coupling-assisted protocol more robust against a temperature increase.

To illustrate the competitivity of these protocols, we consider a  situation where, in a real experimental setup exploiting the feedback protocol, instead of using the feedback loop one hypothetically satisfies the optimal conditions for the dissipative-coupling assisted  protocol.
We take a resent experimental paper~\cite{rossi2018} reporting a record-deep feedback assisted cooling, the experimental parameters of which read
$$
n_{\mathrm{th}}\cong10^5\qquad Q=10^9\qquad \gamma/\omega_M=16\qquad \eta_{\mathrm{det}}=0.77.
$$
This paper also documents the value of $n_{\mathrm{imp}}=5.8\cdot10^{-8}$, which is three orders of magnitude smaller than  previously reported values.
For the laser power used, the estimate (\ref{fbn}) was dominated by  the detector controlled limit $n_{\mathrm{fb}} = 0.07$ while the minimal number of phonon measured experimentally was about $0.3$.

At the same time, for the experimental parameters from this paper, the dissipative-coupling-assisted cooling protocol predicts $n_{\mathrm{diss}}=0.02$ as a cooling limit, which is lower than  $n_{\mathrm{det}}=0.07$ and close to the value of the second term in (\ref{fbn}).
Thus, the dissipative-coupling-assisted protocol looks competitive, if the conditions for its implementation  are met.
One readily checks that the requirement of sufficiently low internal loss Eq.(\ref{crossover1}) is the most demanding.
For the above parameters, via (\ref{gamm/gamm}) and (\ref{ndisp}), it implies
\begin{equation}
\label{gamm0/gamm}
\frac{\gamma_{\mathrm{int}}}{\gamma}\ll \frac{n_{\mathrm{diss}}}{2n_{\mathrm{disp}}}\approx 0.6\cdot10^{-3}.
\end{equation}
Clearly, it is a very demanding requirement, which probably makes it impossible to reach the cooling given by Eq.(\ref{ndiss}) for the system parameters from Ref.~\citenum{rossi2018}.
If this requirement is not met, the cooling limit will be given by Eq.(\ref{n31}) such that the ground-state cooling becomes problematic.
In addition, one should realize that the implementation of the dissipative-coupling-assisted protocol may require an unrealistically high number of in-cavity photons.

\section{Conclusions}
\label{Conclusions}
It was shown that the advanced dissipative-coupling-assisted cooling limit $n_{\mathrm{diss}} $, Eq.(\ref{ndiss}), derived in Ref.~\citenum{Weiss2013a} is valid if it is lower than the dispersive-coupling-assisted limit under the red sideband excitation $n_{\mathrm{disp}} $, Eq.(\ref{ndisp}).
Strictly, the range of applicability of this result is given by Eqs.(\ref{iequ}) and (\ref{iequ1}), which can also be rewritten as follows
\begin{equation}
\label{criterion}
\frac{n_{\mathrm{th}}}{Q}\ll \frac{1}{16}\left(\frac{\gamma}{\omega_M}\right)^3
\qquad \mathrm{and}\qquad
\frac{n_{\mathrm{th}}}{Q}\ll \frac{1}{16}\frac{\gamma}{\omega_M}.
\end{equation}
Otherwise the light-pressure effect makes the  mechanical oscillator overdamped while the weak-coupling regime does not take place such that the theory goes out of its range of applicability and its results do not hold any more.

As expected, the situation with the Fano-effect-driven  cancellation of the otherwise leading contribution results in  stringent requirements from the accuracy of satisfying  the conditions needed to reach the predicted idealized cooling limit.

The internal cavity loss, ignored by the original theory, may affect the cooling limit already when the associated  decay rate $\gamma_{\mathrm{int}}$ is much smaller than the external cavity decay rate $\gamma$:
the internal cavity loss becomes relevant when $\gamma_{\mathrm{int}}$ is about the light-pressure-induced mechanical decay rate, which is much smaller than $\gamma$.
Alternatively, the condition providing to neglect the internal loss can be written as follows
\begin{equation}
\label{gamm0/gamm1}
\frac{\gamma_{\mathrm{int}}}{\gamma}\ll \frac{n_{\mathrm{diss}}}{2n_{\mathrm{disp}}}.
\end{equation}

A similar situation takes place with the accuracy of satisfying the optimized conditions for the detuning and coupling-constant ratio.
Such an inaccuracy may essentially affect the idealized cooling limit already in the regimes where the relative inaccuracy of these parameters is small.

It was also shown that the aforementioned Fano-effect-driven cancelation is lifted in terms of more precise calculations.
As a result, in reality, the idealized cooling limit may be substantially affected.

An instructive conclusion of the paper states that, in the sideband resolved regime where the dispersive-coupling-assisted  protocol is commonly viewed as the ultimate tool, the dissipative-coupling-assisted protocol  may require a much smaller in-cavity photon number for the same cooling level.

The material of the present paper clearly suggests  that  the dissipative-coupling-assisted cooling protocol is competitive once it is perfectly implemented, which, however,  may be challenging.
Here the stringent limitations on the realization of the idealized scenario, which were addressed in this paper, may be essential.
\begin{acknowledgments}
The author acknowledges reading the manuscript by M. Nagoga, G. Avakiants, and M. Olkhovich.
\end{acknowledgments}
\appendix
\section{Stochastic backaction force in Michelson-Sagnac interferometer }
\begin{figure}
\includegraphics [width=0.7\columnwidth,clip=true, trim=0mm 0mm 0mm 0mm] {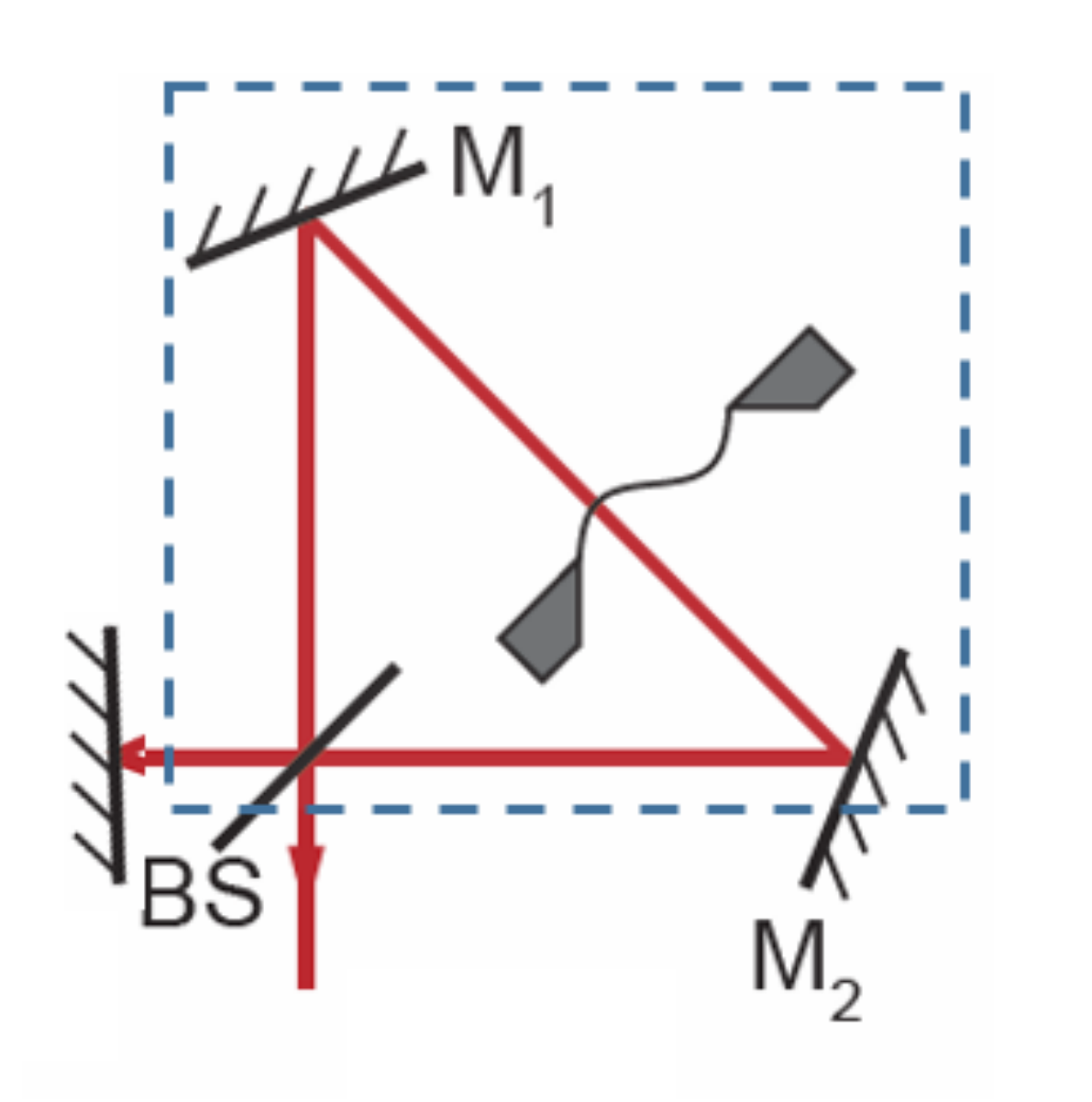}
\caption{Schematic of  Michelson-Sagnac interferometer. The part marked with a dashed-line rectangle  can be considered as an effective input  mirror with $x$-dependent parameters such that the system can be viewed as a one-sided cavity.
\label{MSI5}}
\end{figure}

The Michelson-Sagnac interferometer (MSI) is schematically depicted in Fig.\ref{MSI5}.
In this setup, the beam splitter (BS) and the membrane, shown with a wiggled line, are characterized by following scatting matrices
\begin{equation}
\label{BS}
 \left(
  \begin{array}{cc}
T &  -R \\
R  &  T  \\
  \end{array}
\right)
\qquad \textrm{and}
\qquad
 \left(
  \begin{array}{cc}
-r &  t \\
t  &  r \\
  \end{array}
\right),
\end{equation}
where all coefficients of the matrices are real and positive, and $t$ and $T$ stand for the transmission coefficients.
All mirrors impose a $\pi$ phase shift at reflection.
The membrane is displaced to the left from its symmetric position  by the distance $x$.
The BS-M1 and BS-M2 distances equal $L_a$.
The M1-M2 distance equals $2l$.
The end-mirror-BS distance equals $l_s$.
The part of MSI marked with the dashed rectangle can be considered as an effective mirror.
The whole MSI can be treated as  an optomechanical Fabry-Perrot cavity of a fixed length $L = L_a+l+l_s$ with the input mirror, the scattering matrix of which reads~\cite{Tarabrin2013}
\begin{equation}
\label{Matrix}
\mathbb{M}= \left(
  \begin{array}{cc}
\rho & \tau \\
\tau &-\rho^* \\
  \end{array}
\right),\qquad \rho=|\rho| e^{i\mu}
\end{equation}
\begin{equation}
\label{MSrho}
\rho=  -2RTt -(R^2-T^2) r \cos2kx + ir \sin2kx,
\end{equation}
\begin{equation}
\label{MStau}
\tau= t(T^2-R^2)+2RTr \cos2kx,
\end{equation}
where $\tau$ stands for the transmission coefficient.
Equations (\ref{MSrho}) and (\ref{MStau}) are written for a wave with  wave vector $k$.
The interferometer decay rate $\gamma$ and resonance frequencies $\omega_c$ can be written as
\begin{equation}
\label{MSgamma}
\gamma= \frac{\tau^2c}{2L}.
\end{equation}
\begin{equation}
\label{omega}
\omega_c= \frac{c}{2L}(2\pi N -\mu)
\end{equation}
where $N$ is integer and $c$ is the light velocity.

Since at resonance  $\omega_c=ck$, in view of a $k$ dependence of $\mu$,  (\ref{omega}) is  an equation for $\omega_c$.
However, if the membrane displacement $x$ is much smaller than $L$, the dispersive coupling constant can be calculated neglecting the $k$-dependence of $\mu$ to find
\begin{equation}
\label{go}
g_\omega=-\frac{d\omega_c}{dx}x_{\textrm{zpf}} =\frac{d\mu}{dx}
\frac{ c}{2L}x_{\textrm{zpf}}.
\end{equation}
\begin{equation}
\label{gg}
g_\gamma=-\frac{1}{2}\frac{d\gamma}{dx}x_{\textrm{zpf}}
=-\tau\frac{d\tau}{dx}\frac{ c}{2{L}}x_{\textrm{zpf}}
\end{equation}
where
\begin{equation}
\label{derivatives}
  \begin{array}{cc}
&\frac{d \tau}{d x}= -4krRT\sin2kx  \\
&\frac{d \mu}{d x}= -2kr[2tRT\cos2kx-r(T^2-R^2)].  \\
  \end{array}
\end{equation}

Reference~\cite{Tarabrin2013} addresses the linear optomechanics of such an interferometer when it is under a strong monochromatic excitation with a frequency $\omega_L$.
In our notation, the spectral power density calculated for the stochastic backaction force acting on the membrane reads
\begin{equation}
\label{SFTAR}
S_{FF}(\omega) = \left(\frac{\hbar\omega_L|a_0|}{L}\right)^2\frac{r}{\gamma} \frac{|N(\omega)|^2}{|1-e^{2i(\omega_L+\omega)L/c+i\mu}|^2}.
\end{equation}
\begin{equation}
\label{N2}
N(\omega)= \alpha_1(1+e^{2iL\omega/c})+\alpha_2e^{2ikL}  +\alpha_2^*e^{-2iL\omega_L/c}
\end{equation}
\begin{equation}
\label{alpha1}
\alpha_1= 2tRT \cos2kx- r(T^2-R^2)
\end{equation}
\begin{equation}
\label{alpha2}
\alpha_2 = \cos2kx +i(T^2-R^2)\sin2kx,
\end{equation}
We are interested in the lowest order terms in $\omega=ck-\omega_L$, detuning $\Delta=\omega_L-\omega_c$, and $|\tau|$.

Thus, keeping in mind the resonance condition
\begin{equation}
\label{Exp1}
e^{2iL\omega_c/c+i\mu}=1,
\end{equation}
we approximate
\begin{equation}
\label{Exp2}
e^{2iLk_L/c}\approx e^{-i\mu}(1+2i\Delta L/c)\qquad e^{2ikL}\approx  e^{-i\mu}\left[ 1+2i(\Delta+\omega)L/c\right]
\end{equation}
to present (\ref{N2}) as
\begin{equation}
\label{N3}
N(\omega)= 2(\alpha_1+\textrm{Re}[\tilde{\alpha}_2])(1+iL\omega/c)
-2\textrm{Im}[\tilde{\alpha}_2](2\Delta+\omega)L/c \qquad \tilde{\alpha}_2=e^{-i\mu}\alpha_2.
\end{equation}
Next, taking into account that, in the accepted approximation
\begin{equation}
\label{alpha11}
\alpha_1 = -\frac{c|\rho|^2}{2\omega_L r}\frac{\partial \mu}{\partial x}\qquad \tilde{\alpha}_2 =-\frac{\alpha_1}{|\rho|}+ i\frac{c}{2\omega_L r|\rho|} \tau\frac{\partial \tau}{\partial x},
\end{equation}
we can write
\begin{equation}
\label{N4}
N(\omega)= \left( \frac{2L}{c}\right)^2\frac{1}{x_{\textrm{zpf}}}\frac{c\gamma}{2\omega_L r}
\left[g_\omega (1+iL\omega/c)+g_\gamma\frac{2\Delta+\omega}{\gamma}\right]
\end{equation}

Finally, Eqs.(\ref{SFTAR}) and (\ref{N4}) bring us to Eq.(48) from the main text.
\bibliography{QOwork,NF}

\begin{thebibliography}{32}
\bibitem [{foo({\natexlab{b}})}]{footnote5}%
  \BibitemOpen
  \href@noop {} {\bibfield  {journal} {\bibinfo  {journal} {There exists a
  confusion with the term "input-output". In the Langevin equation method, the
  realtions linking the operators of the fields inside and outside the cavity
  are also called input-output realations}\ } ({\natexlab{b}})}\BibitemShut
  {NoStop}%
\end{thebibliography}

\end{document}